\documentclass[footinbib,aps,pra,reprint,superscriptaddress,nopacs]{revtex4-1}

\usepackage{amsmath,amssymb,bbm,bm,graphicx}
\usepackage{csquotes}
\usepackage{amsfonts}
\usepackage{verbatim}
\usepackage{braket}   
\usepackage{mathtools}

\usepackage{amsthm,mathrsfs,amsfonts,epsfig,bm,mathrsfs,enumerate,algorithm,algpseudocode}
\usepackage[shortlabels]{enumitem}

\def\bx{\mathbf{x}}

\def\bbD{{\bm{\mathcal{D}}}}
\def\cN{\mathcal{N}}
\DeclareMathOperator*{\argmin}{argmin}
\DeclareMathOperator{\Tr}{Tr}
\DeclareMathOperator{\R}{Re}
\DeclareMathOperator{\I}{Im}

\usepackage[absolute]{textpos}
\usepackage{xcolor}

\begin{document}
\title{A Bayesian analysis of classical shadows}

\author{Joseph M. Lukens}
\email{lukensjm@ornl.gov}
\affiliation{Quantum Information Science Group, Oak Ridge National Laboratory, Oak Ridge, Tennessee 37831, USA}

\author{Kody J. H. Law}
\affiliation{School of Mathematics, University of Manchester, Manchester, M13 9PL, UK}

\author{Ryan S. Bennink}
\affiliation{Quantum Computational Science Group, Oak Ridge National Laboratory, Oak Ridge, Tennessee 37831, USA}

\date{\today}

\begin{abstract}
The method of classical shadows heralds unprecedented opportunities for quantum estimation with limited measurements [H.-Y. Huang, R. Kueng, and J. Preskill, Nat. Phys. \textbf{16}, 1050 (2020)]. Yet its relationship to established quantum tomographic approaches, particularly those based on likelihood models, remains unclear. In this article, we investigate classical shadows through the lens of Bayesian mean estimation (BME). In direct tests on numerical data, BME is found to attain significantly lower error on average, but classical shadows prove remarkably more accurate in specific situations---such as high-fidelity ground truth states---which are improbable in a fully uniform Hilbert space. We then introduce an observable-oriented pseudo-likelihood that successfully emulates the dimension-independence and state-specific optimality of classical shadows, but within a Bayesian framework that ensures only physical states. Our research reveals how classical shadows effect important departures from conventional thinking in quantum state estimation, as well as the utility of Bayesian methods for uncovering and formalizing statistical assumptions.
\end{abstract}

\maketitle
\section*{Introduction}
\label{sec:intro}
Measurement and characterization of quantum systems comprise a long-standing problem in quantum information science~\cite{James2001}.  However, the exponential scaling of Hilbert space dimension with the number of qubits makes full characterization extremely challenging, inspiring a plethora of approaches designed to estimate properties of quantum states with as few measurements as possible, such as compressed sensing~\cite{Gross2010, Flammia2012}, adaptive tomography~\cite{Huszar2012, Kravtsov2013, Granade2017}, matrix product state formulations~\cite{Cramer2010}, and neural networks~\cite{Torlai2018, Carrasquilla2019, Lohani2020}. Very recently, a groundbreaking approach known as classical shadows was proposed and analyzed~\cite{Huang2020}. Building on and simplifying ideas from ``shadow tomography''~\cite{Aaronson2018}, the classical shadow was shown to provide accurate predictions of observables with a fixed number of measurements, including simulated examples for quantum systems in excess of 100 qubits~\cite{Huang2020}. Astonishingly simple, the classical shadow is formed by collecting the results of random measurements on a repeatedly prepared input state, and inverting them through an appropriate virtual quantum channel.

However, several features of the classical shadow remain enigmatic, including its highly nonphysical nature, optimality with respect to alternative cost functions, and  relationship to more conventional likelihood-based tomographic techniques. One such method, Bayesian mean estimation (BME)~\cite{Blume2010}, provides a conceptually straightforward path to estimate a quantum state given measured data, making use of prior knowledge and providing meaningful error bars for any experimental conditions. BME appears particularly well suited for contextualizing classical shadows, since it returns a principled estimate under any number of measurements (even zero), and is optimal in terms of minimizing average squared error~\cite{Robert1999}.

In this work, we directly compare the estimates of classical shadows and BME for identical simulated datasets. For particular observables with relatively improbable values from the perspective of BME, shadow is found to reach the ground truth with significantly fewer measurements. However, after properly reformulating the problem under test for consistency with the Bayesian prior, the situation reverses, with BME returning estimates possessing lower error on average. In the latter portion of our investigation, we seek to construct a BME model emulating the key features of the classical shadow, but with positive semidefinite states as support. While complicated by the shadow's nonphysical nature, we ultimately propose an observable-oriented pseudo-likelihood that rates quantum states by their observable values with respect to those of shadow. Our pseudo-likelihood successfully mimics the dimension-independence of shadow, with the advantage of delivering entirely physical estimates for any number of measurements.

\section*{Results}
\label{sec:res}

\subsection*{Problem Formulation}

\textit{Classical Shadows.} For our analysis, we invoke the setup of the original classical shadow proposal~\cite{Huang2020}. Consider a $D$-dimensional Hilbert space occupied by a ground truth quantum state $\rho_g$ that can be repeatedly prepared. On each preparation $m$, $\rho_g$ is subjected to a randomly chosen $D\times D$ unitary $U_m$ and one measurement is performed in the computational basis, leaving result $\ket{b_m}$. Defining $\ket{\psi_m} = U_m^\dagger\ket{b_m}$, the classical snapshot associated with measurement $m$ follows as $\mathcal{M}^{-1}(\ket{\psi_m}\bra{\psi_m})$, where $\mathcal{M}(\cdot)$ is the quantum channel defined by averaging over all possible unitaries and outcomes.

We assume the $U_m$ are drawn from the set of $D\times D$ Haar-random unitaries, in which case $\mathcal{M}^{-1}(\ket{\psi_m}\bra{\psi_m}) = (D+1)\ket{\psi_m}\bra{\psi_m} - I_D$, with $I_D$ the $D\times D$ identity matrix~\cite{Huang2020}. (This channel holds for the more restricted class of random Cliffords as well~\cite{Webb2016, Zhu2017}.) Averaging over $M$ measurements yields the shadow estimator
\begin{equation}
\label{eq:shad}
\rho_s = \frac{D+1}{M} \sum_{m=1}^M \ket{\psi_m}\bra{\psi_m} - I_D.
\end{equation}
(In what follows, the phrases ``classical shadow,'' ``shadow estimator,'' and simply ``shadow'' refer interchangeably to this estimator as well as the procedure more generally.) In this form, the simplicity of $\rho_s$ is evident: it is merely a scaled and recentered average of all observed outcomes. Interestingly, though, $\rho_s$ is in general not positive semidefinite; for $M<D$, $\rho_s$ possesses at least $D-M$ eigenvalues equal to $-1$. Accordingly, in the targeted regime for classical shadows of $M\ll D$, $\rho_s$ is highly nonphysical.
Understanding the role the shadow estimator's negativity on estimation forms a central goal of the present study. Finally, defining $\lambda$ as the expectation of the observable $\Lambda$ ($\lambda = \Tr\rho\Lambda$), the shadow estimate thereof follows as 
\begin{equation}
\label{eq:s}
\lambda^{(s)} = \Tr\rho_s\Lambda,
\end{equation}
to be compared to the ground truth $\lambda^{(g)} = \Tr\rho_g\Lambda$.

As an aside, we note that Ref.~\cite{Huang2020} employed an additional statistical technique, ``median of means,'' to reduce the impact of outliers by partitioning the $M$ outcomes into $K$ subsets and taking the median as the estimate $\lambda^{(s)}$.  In the interests of simplicity and ease of comparison, we focus on $K=1$ in Eq.~(\ref{eq:shad}). We expect the benefits of selecting $K>1$ will prove similar in both the shadow and Bayesian cases~\cite{Orenstein2019}, but work on this is beyond the scope of the present investigation.

\textit{Bayesian Mean Estimation.} In the Bayesian paradigm, the same set of measurement outcomes $\bbD = \{\ket{\psi_1}, \ket{\psi_2},...,\ket{\psi_M} \}$ is related to a possible density matrix $\rho(\bx)$ via a likelihood consisting of the product of probabilities set by Born's rule:
\begin{equation}
\label{eq:LL}
L_\bbD(\bx) = \prod_{m=1}^M \braket{\psi_m|\rho(\bx)|\psi_m},
\end{equation}
that is, $L_\bbD(\bx)\propto \Pr(\bbD|\rho)$---the probability of receiving the dataset $\bbD$ given quantum state $\rho$. Some prior distribution $\pi_0(\bx)$ is also assumed, defined for parameters $\bx$ such that $\rho(\bx)$ is always physical: trace-one, Hermitian, and positive semidefinite. Then the posterior describing the distribution of $\bx$ given the observed data $\bbD$ ensues from Bayes' rule:
\begin{equation}
\label{eq:post}
\pi(\bx) = \frac{1}{\mathcal{Z}} L_\bbD(\bx) \pi_0 (\bx).
\end{equation}
Note that the randomness of the chosen unitaries $U_m$ does not enter the Bayesian model; only the outcomes $\ket{\psi_m}$ play a role. The selection of unitary $U_m$ is independent of the (unknown) density matrix, i.e., $\Pr(U_m=U|\rho) = \Pr(U_m=U)$; thus any probabilities would cancel out through the normalization factor $\mathcal{Z}$ in Eq.~(\ref{eq:post}). Intuitively, in the Bayesian view the experimenter knows the unitaries exactly post-experiment, regardless of how they were chosen, so imposing uncertainty on them in the estimation process proves superfluous. Consequently, while the uncertainty of BME depends strongly on the variety of measurements chosen, the theory does not, a conspicuous departure from shadow where the distribution of $U_m$ enters directly through the inverted quantum channel $\mathcal{M}^{-1}(\cdot)$. 

Formally, the posterior distribution in Eq.~(\ref{eq:post}) completes the Bayesian model. From this, one can estimate any function of $\rho(\bx)$. For the most direct comparison with the classical shadow, here we focus on BME specifically, which for some observable $\Lambda$ is the point estimate defined as
\begin{equation}
\label{eq:B}
\begin{split}
\lambda^{(B)} & = \braket{\Tr \rho\Lambda}_\rho \\
 & = \int d\bx\, \pi(\bx) \, \Tr \rho(\bx)\Lambda \\
 & = \Tr\left\{ \left[ \int d\bx\, \pi(\bx) \rho(\bx) \right] \Lambda \right\} \\
 & = \Tr \rho_B\Lambda,
\end{split}
\end{equation}
where the last two lines follow, respectively, from the linearity of the trace operation and defining the Bayesian mean $\rho_B = \int d\bx\, \pi(\bx) \rho(\bx)$. This convenient simplification, in which the Bayesian mean of a quantity is simply its value at $\rho_B$, holds for linear functions of $\rho$, which includes all quantum observables and which we focus on in this article. Moreover, $\lambda^{(B)}$ is the function of $\bbD$ which minimizes the mean-squared error (MSE) averaged over all possible states and outcomes. That is,
\begin{equation}
\label{eq:opt}
\lambda^{(B)} = \argmin_{\lambda(\bbD)} \int d\bbD \int d\bx\, \pi(\bx,\bbD) \left[ \lambda(\bbD) - \Tr \rho(\bx)\Lambda \right]^2,
\end{equation}
with $\pi(\bx,\bbD)$ the joint distribution over data and parameters~\cite{Robert1999}. This optimality is nonasymptotic, holding for any number or collection of unitaries $\{U_1, U_2,..., U_M\}$. Considering the widely different expressions for $\rho_s$ [Eq.~(\ref{eq:shad})] and $\rho_B$ [Eq.~(\ref{eq:B})], we found it remarkable just how well $\rho_s$ performed in Ref.~\cite{Huang2020} in light of BME's optimality in Eq.~(\ref{eq:opt}); it was this feature which initially inspired us to develop a thorough comparison between shadow and BME.

\textit{Simulated Experiments}
In general, comparing the performance of estimators derived from classical (frequentist) statistics---like $\rho_s$---with those from Bayesian methods proves tricky business, since they view uncertainty in functionally different ways. Therefore we adopt a pragmatic view which aligns with the interests of experimentalists: perform experiments, compute the associated shadow and BME estimators, and calculate their error with respect to actual values. While the final step is not always possible in practice, it is in numerical simulation, where the ground truth $\rho_g$ is known exactly. Doing so enables us to illuminate the advantages and disadvantages of both approaches on equal footing. We employ the approach described in the ``Methods'' section for obtaining simulated datasets $\bbD$.

\begin{figure*}[tb!]
\centering\includegraphics[width=2\columnwidth]{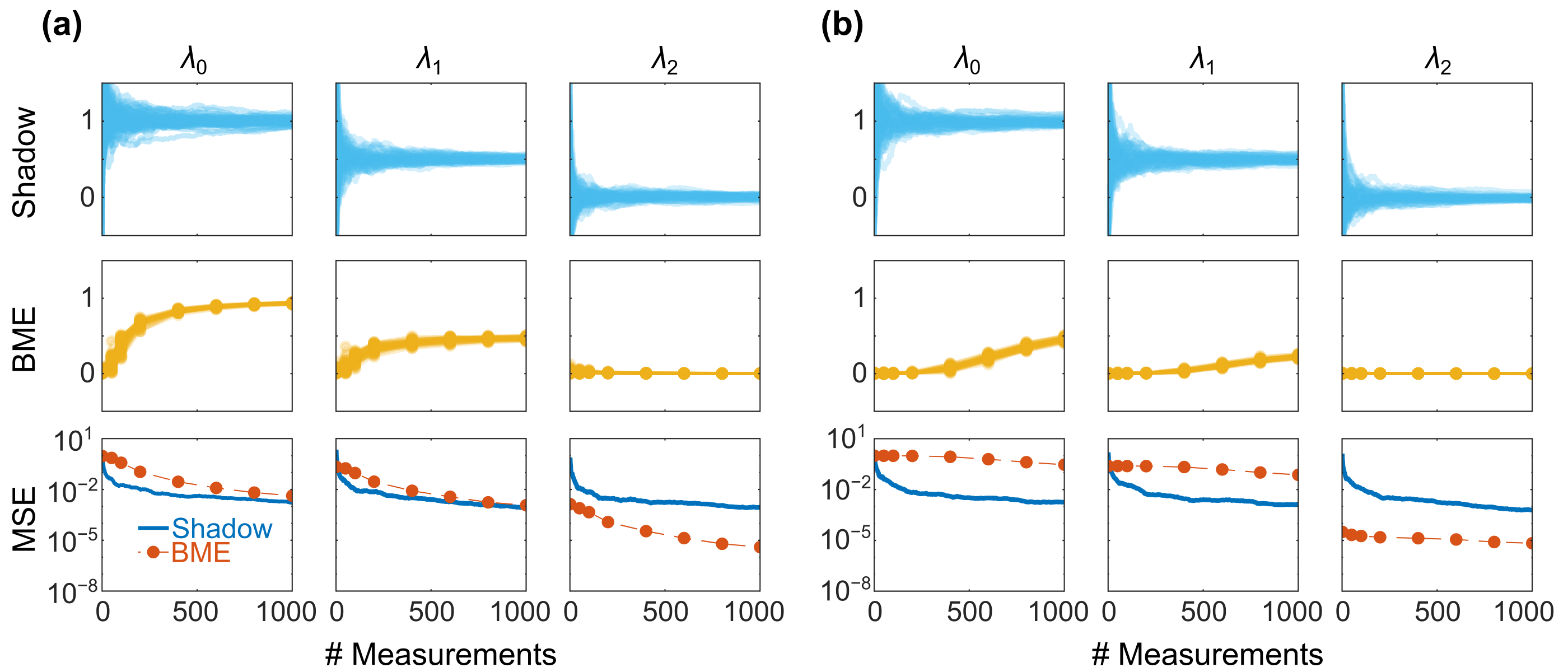}
\caption{Comparison of shadow and BME estimates of $\lambda_n$ for (a) $D=32$ and (b) $D=256$. Results from fifty trials for each dimension are plotted, assuming a fixed ground truth state $\ket{0}$.}
\label{fig1}
\end{figure*}

\subsection*{Comparing Classical Shadows and BME}
\label{sec:compare}

\textit{Picture 1: Fixed Ground Truth.} 
As our first benchmark, we compare the performance of $\rho_s$ and $\rho_B$ in estimating three rank-1 observables, of which fidelities and entanglement witnesses form an important and experimentally relevant subset. Specifically, we consider $\Lambda_n = \ket{\phi_n}\bra{\phi_n}$ ($n=0,1,2$) where
\begin{equation}
\label{eq:states}
\begin{split}
\ket{\phi_0} & = \ket{0}\\
\ket{\phi_1} & = \frac{1}{\sqrt{2}}\ket{0} + \frac{1}{\sqrt{2(D-1)}}\sum_{j=1}^{D-1} \ket{j}\\
\ket{\phi_2} & = \ket{1}.
\end{split}
\end{equation}
These possess ground truth values equally spaced within the physically allowed range for trace-one, rank-one observables: $\lambda_0^{(g)}=1$, $\lambda_1^{(g)}=\frac{1}{2}$, and $\lambda_2^{(g)}=0$. The shadow estimator is readily obtained from Eq.~(\ref{eq:shad}), so we compute $\rho_s$ for all $M\in\{1,2,...,1000\}$, where $M$ defines the set containing the first $M$ measurements: $\bbD=\{\ket{\psi_m};\; m=1,2,...,M\}$.

On the other hand, $\rho_B$ requires evaluation of the high-dimensional integral $\int d\bx\,\pi(\bx)\rho(\bx)$. To that end, we summon Markov chain Monte Carlo (MCMC) methods, several of which have been explored in the context of quantum state estimation, including Metropolis--Hastings~\cite{Blume2010, Mai2017}, Hamiltonian Monte Carlo~\cite{Seah2015}, sequential Monte Carlo (SMC)~\cite{Granade2016}, and slice sampling~\cite{Williams2017, Lu2019a}. We select the preconditioned Crank--Nicolson algorithm~\cite{Cotter2013} applied in Ref.~\cite{Lukens2020}, which to our knowledge is the most efficient BME approach currently available for density matrix recovery. Finally, because of our assumed pure state ground truth, we take as prior all pure states $\rho=\ket{\psi}\bra{\psi}$ uniformly distributed on the complex $D$-dimensional unit hypersphere. Numerically, the parameters $\bx$ reduce to a $D$-dimensional complex column vector, so we have $\pi_0(\bx)\propto \exp\left(-\frac{1}{2}\bx^\dagger\bx\right)$, $\rho(\bx) = \frac{\bx\bx^\dagger}{|\bx|^2}$, and $d\bx = \prod_{l=1}^{D} d(\R x_l) d(\I x_l)$ with $x_l$ denoting a single component of $\bx$.

The use of pure states is not central to the BME formalism whatsoever, but does permit us to simulate in higher dimensions than otherwise possible. With pure states only, our parameterization entails $2D$ real numbers, compared to $2D^2+D$ for mixed states. As an example, for $D=256$, the pure state prior, and likelihood of Eq.~(\ref{eq:LL}), each MCMC chain takes about ten minutes to converge on our desktop computer, which for the 400 settings involved in Fig.~\ref{fig1}(b) amounts to $\sim$2.5 days. Based on previous studies~\cite{Lukens2020} the mixed state version would therefore have been completely unfeasible at this dimension with our computational resources, likely taking weeks (or more) to complete~\footnote{Incorporating some of the methods suggested in Ref.~\cite{Lukens2020} in further research, such as embedding within SMC samplers and parallelization, should permit the extension to significantly larger $D$ and mixed states}. With pure states, then, we can focus more directly on dimensional scaling and the statistics from many trials.

For each trial, we perform BME for eight collections of measurements $M\in\{1,50,100,200,400,600,800,1000\}$. We keep $R=2^{10}$ samples from each chain of length $RT$, where we select the thinning factor $T$ empirically to obtain convergence. Figure~\ref{fig1} plots the estimates for all 50 trials obtained by both shadow and BME for $D=32$ [Fig.~\ref{fig1}(a)] and $D=256$ [Fig.~\ref{fig1}(b)]. A thinning value of $T=2^9$ ($T=2^{12}$) is used for $D=32$ ($D=256$). Each column corresponds to a particular expectation value $\lambda_n$; the bottom row shows the MSE with respect to the ground truth, averaged over all trials defined as $\braket{|\lambda_n^{(\cdot)} - \lambda_n^{(g)}|^2}_\mathrm{trials}$ with $\cdot=s$ for the shadow and $\cdot=B$ for BME. The classical shadows show wide variation for low $M$, including highly nonphysical estimates ($\lambda_n^{(s)}>1$ or $<0$), but they converge to ground truth values rapidly, with nearly identical rates for all observables and dimensions. This is confirmed quantitatively in the MSE curves that attain values of $\sim$10$^{-3}$ by $M=1000$ for all cases.

The behavior proves vastly different for BME. While physical estimates are always returned, the number of measurements needed to reach the ground truth varies strongly both with observable $\lambda_n$ and with dimension $D$. Intriguingly, shadow shows significantly lower MSE for $\lambda_0$ and $\lambda_1$, widening as $D$ increases. On first glance, this presents a paradox: Eq.~(\ref{eq:opt}) implies that $\lambda_n^{(B)}$ should possess the lowest possible MSE for any $n$ and $M$, and yet $\lambda_n^{(s)}$ convincingly surpasses it these cases. Yet this dilemma can be resolved by studying the prior $\pi_0(\bx)$. When the Bayesian model assigns equal \emph{a priori} weights to all possible states---a sensible choice for an uninformative prior---this by implication makes observable values such as $\lambda_0^{(g)}=1$ highly unlikely, since only one state in the domain attains this. On the other hand, expectations for any rank-1 projector $\Lambda$ on the order of $\lambda\sim \frac{1}{D}$ are to be expected initially since $\int d\bx\,\pi_0(\bx) \Tr \rho(\bx)\Lambda = \frac{1}{D}$. This manifests itself in Fig.~\ref{fig1} in BME's much lower MSE for $\lambda_2$, whose ground truth value $\lambda_2^{(g)}=0$ is much more probable. Thus, by running 50 repeated trials with the \emph{same ground truth} $\rho_g=\ket{0}\bra{0}$, the situation over which we average does not accurately reflect the uninformative prior; the conditions for BME optimality are not met.

\textit{Picture 2: Random Ground Truth.} To accurately reflect uninformative prior knowledge, we therefore must prepare \emph{random} ground truth states in our simulations. To do so, we leverage the equivalence between (i) randomly prepared input states with a fixed observable---the situation of interest---and (ii) random selection of an observable for a fixed input. Consider the expectation of observable $\Lambda$, where the quantum state is rotated by some random unitary $U$:
\begin{equation}
\label{eq:equiv}
\Tr \left[(U\rho U^\dagger)\Lambda\right]= \Tr \left[\rho(U^\dagger\Lambda U)\right].
\end{equation}
Thus one can emulate the effect of a randomized state by randomly rotating the observable and evaluating it on a fixed state. Practically speaking, we are free to employ the same simulated datasets and estimators $\rho_s$ and $\rho_B$ above, but select at random a different projector $\Lambda=\ket{\phi}\bra{\phi}$ for each trial. This is equivalent to performing all trials with a random ground truth but a fixed observable. We call this randomized evaluation ``Picture 2'' to distinguish it from the fixed ground truth case above (Picture 1).

\begin{figure*}[tb!]
\centering\includegraphics[width=2\columnwidth]{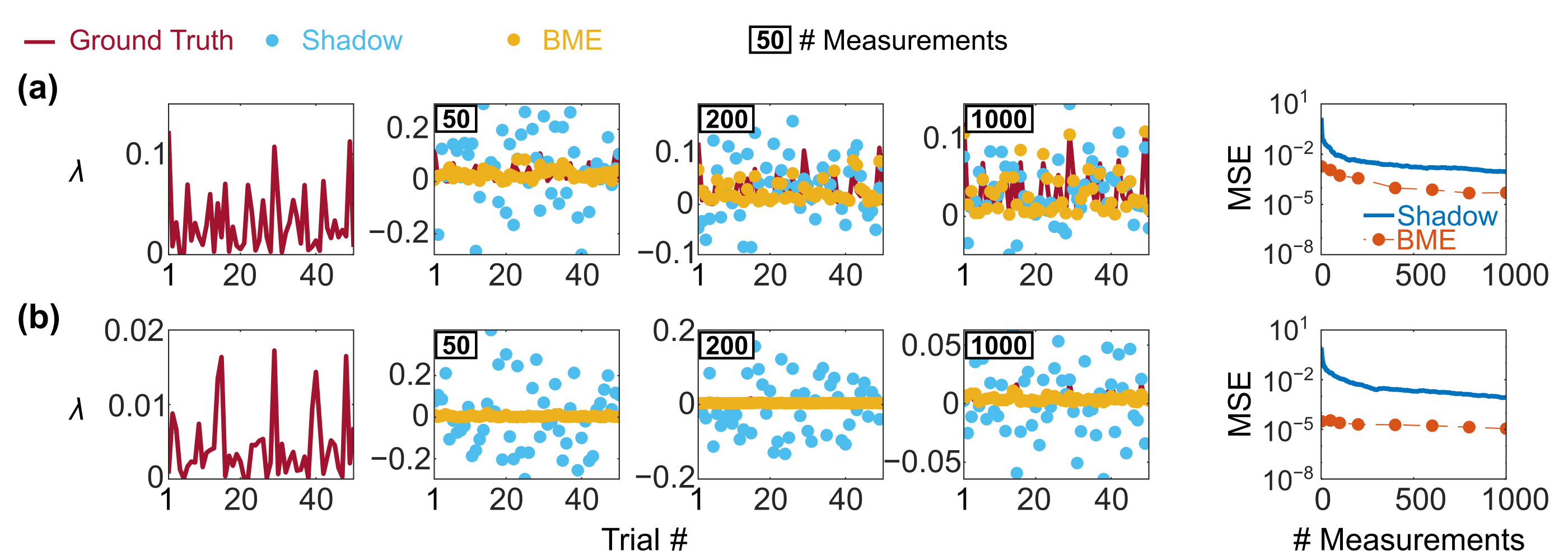}
\caption{Estimating rank-1 observable $\Lambda$ for randomly chosen ground truth states (Picture 2). (a) $D=32$ case. (b) $D=256$ case. The first four columns show $\lambda$ values for each trial; the last column plots MSE with respect to ground truth over all trials.}
\label{fig2}
\end{figure*}

Results appear in Fig.~\ref{fig2} for (a) $D=32$ and (b) $D=256$. The first column plots the ground truth value $\lambda^{(g)}$ for each trial, the next three columns plot the shadow and BME estimates for increasing numbers of measurements, and the final column presents the MSE with respect to the ground truth. 
Now BME returns much more accurate estimates than shadow on average, and the paradox regarding Bayesian optimality is solved: the Bayesian mean gives the lowest MSE as long as the prior accurately reflects the true uncertainty of the system under test. Accordingly, this BME study clarifies an underlying assumption in selecting observables in Picture 1: being able to ``guess'' an observable with such high overlap to the ground truth suggests that one is not really operating under the neutrality implied by a uniform prior; an informative prior would more accurately reflect the situation.

This observation brings to light an interesting question of motivation in a given quantum experiment. In the sense of ensuring that any estimate is adequately justified by the data, the idea of ``baking in'' a prior favoring some subset of quantum states is undesirable. And yet, in many situations the researcher \emph{does} have strong beliefs---or at least hopefulness---about the state being prepared, and wants to verify this by computing an observable, such as fidelity, where it is desired that $\lambda^{(g)} \sim 1$. In this case, one wishes to validate such high values quickly with few measurements, but likely does not care so much about how well the procedure can estimate the ground truth when it is \emph{low} (e.g., when $\lambda^{(g)}\sim \frac{1}{D}$), since this situation suggests a poorly prepared state anyway. Accordingly, the felt cost is stronger when error is higher for situations with $\lambda^{(g)} \gg \frac{1}{D}$ than when $\lambda^{(g)} \sim \frac{1}{D}$, which is not captured by the standard MSE as expressed in Eq.~(\ref{eq:opt}). And as shown in our tests here, it is precisely these improbable situations wherein shadow excels over BME. Thus our simulations reveal one surprising reason classical shadows are so powerful: they perform well within those subspaces of the entire Hilbert space which are of interest to a high-fidelity system.

\begin{figure*}[tb!]
\centering\includegraphics[width=2\columnwidth]{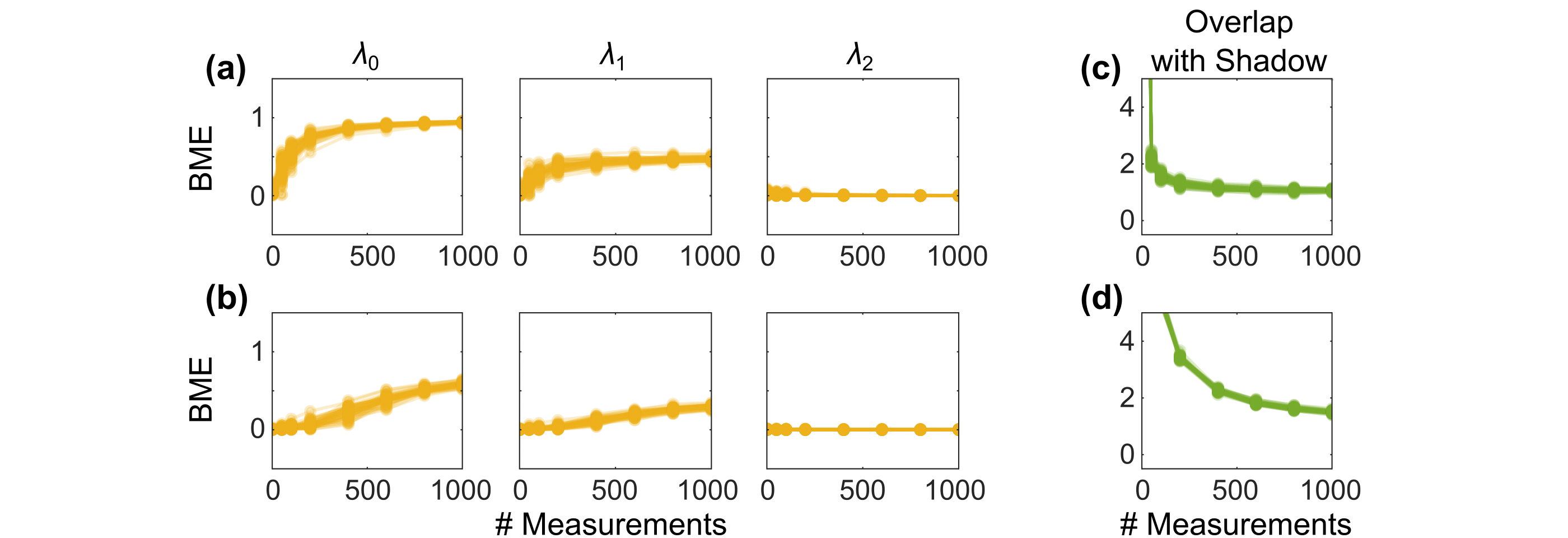}
\caption{Bayesian inference results utilizing the pseudo-likelihood in Eq.~(\ref{eq:PL1}) for (a) $D=32$ and (b) $D=256$. The overlap with shadow, $\Tr\rho_B\rho_s$ is plotted in (c) for $D=32$ and (d) for $D=256$.}
\label{fig3}
\end{figure*}

\subsection*{Emulating Classical Shadows with BME}
\label{sec:emulate}

\textit{Pseudo-Likelihood Formulation.} The dimension-independence and rapid convergence of classical shadows for cases of interest indicate the value of a Bayesian version with similar features, both to gain further insight into shadow itself and to improve thereon by ensuring only physically acceptable states. A simple approach for custom Bayesian models, gaining traction in ``probably approximately correct'' (PAC) learning~\cite{Guedj2019}, proposes use of a pseudo-likelihood that rates a prospective state's suitability through a cost function, instead of a full likelihood based on a physical model. In quantum state tomography in particular, quadratic costs of the form $\lVert \rho-\tilde{\rho} \rVert_F^2$ have been explored~\cite{Mai2017, Lukens2020}, where $\tilde{\rho}$ signifies some point estimator and $\lVert A \rVert_F=\sqrt{\Tr A^\dagger A}$ the Frobenius norm. Therefore, to obtain a physical state with properties similar to $\rho_s$, we first suggest the pseudo-likelihood
\begin{equation}
\label{eq:PL1}
L_\bbD(\bx) = \exp\left(- \frac{K}{2} \lVert \rho(\bx)-\rho_s \rVert_F^2\right).
\end{equation}
The constant $K$ establishes the relative weight of prior and likelihood. Previously, we suggested $K=M$ for reasonable uncertainty quantification~\cite{Lukens2020}; here we consider $K=MD$ to impart dimension-independence. (Incidentally, we have found no significant modifications to the results below when testing with $K\gg MD$.)

Figure~\ref{fig3}(a) and (b) show the BME results obtained for $D=32$ and $D=256$, respectively, where we again return to Picture 1 with fixed ground truth for all trials. For the tests here, thinning of $T=2^8$ ($T=2^{10}$) is used for the $D=32$ ($D=256$) MCMC chains. Compared to the shadow results of Fig.~\ref{fig1}, the BME predictions still do not reach ground truth values for $\lambda_0$ and $\lambda_1$ efficiently. This proves intriguing, since $\lVert \rho-\rho_s \rVert_F^2$ with $\rho=\ket{\psi}\bra{\psi}$ is minimized precisely by states for which $\braket{\psi|\rho_s|\psi}$ is large. So if $\lambda_0^{(s)}=\braket{g|\rho_s|g}\sim 1$ (cf. Fig.~\ref{fig1}), it is odd that predictions using a BME value maximizing $\braket{\psi|\rho_s|\psi}$ looks so different for $D=256$. The origin of this discrepancy, however, lies in $\rho_s$'s nonphysicality.

\begin{figure*}[tb!]
\centering\includegraphics[width=2\columnwidth]{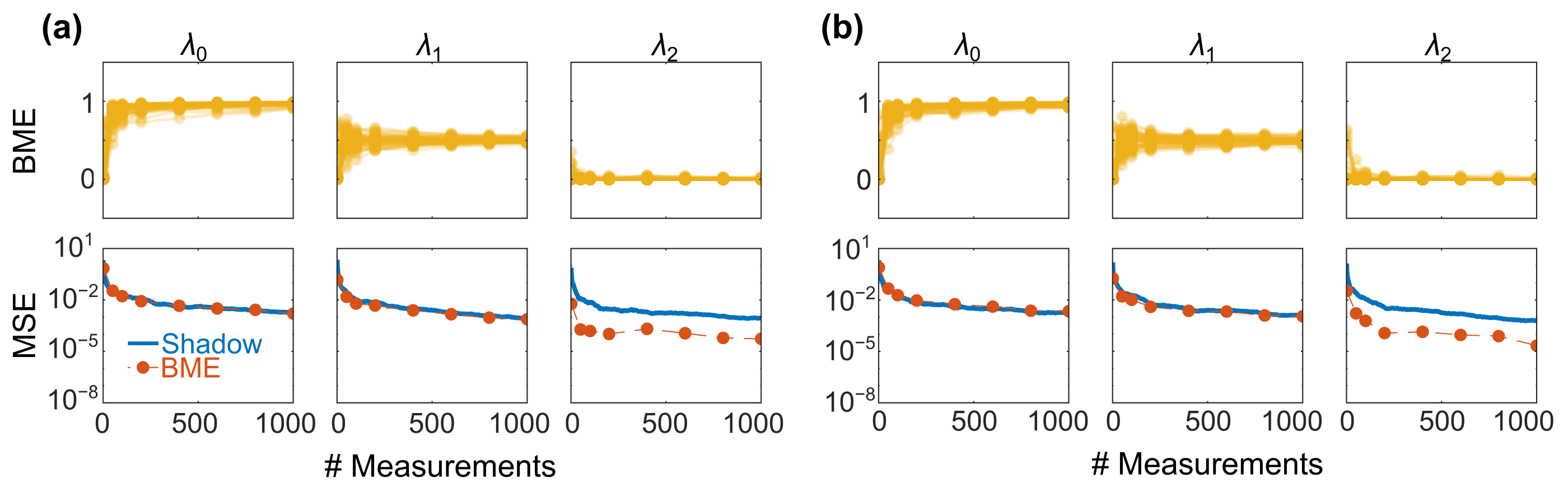}
\caption{Bayesian estimation using the pseudo-likelihood of Eq.~(\ref{eq:PL2}) with $N=3$. (a) Results for $D=32$. (b) Results for $D=256$. The MSE values for shadow from Fig.~\ref{fig1} are reproduced for comparison.}
\label{fig4}
\end{figure*}

Plotting the average overlap between shadow and Bayesian samples ($\Tr\rho_B\rho_s$) in Fig.~\ref{fig3}(c) and (d), we find that $\rho_B$ overlaps with $\rho_s$ \emph{more strongly than the ground truth} $\rho_s=\ket{g}\bra{g}$. Because $\rho_s$ is not positive semidefinite, $\Tr\rho_B\rho_s > 1$ for all cases examined. Thus the BME procedure succeeds in finding states with strong overlap to the shadow, but the closest physical state to $\rho_s$ is not the ground truth, even though $\braket{g|\rho_s|g}\sim 1$.  Intuitively, this nonphysicality helps explain why observables with highly improbable values from the Bayesian view are estimated so much more efficiently with shadow. For a parameterization over physical states and rank-1 observable $\Lambda$, only a single state in the Hilbert space attains $\lambda=1$, and since this represents the maximum value possible for any valid quantum state, it can only be approached from below. On the other hand, a continuum of shadow estimators $\rho_s$ permit $\lambda=1$, for $\rho_s$ is constrained only by Hermiticity and unit-trace---not positive semidefiniteness. Therefore the estimate $\lambda^{(s)}$ can err on either the high or low side (cf. Fig.~\ref{fig1}), pulling the shadow more rapidly to the ground truth in these extreme cases. 

This discloses the second central finding of our investigation: the nonphysicality of $\rho_s$ is not a deficiency, but rather critical to obtaining dimension independence. Thus the key features of the shadow are not necessarily translated onto physical projections like the BME model here~\footnote{As an additional check, we performed the algorithm of Ref.~\cite{Smolin2012} to determine the closest physical density matrix to $\rho_s$, finding very similar results as Fig.~\ref{fig3}. This indicates that our projection conclusions are not an artefact of the pure state prior, but hold for general mixed states as well.} or, for that matter, alternative projected-least-squares approaches~\cite{Smolin2012, Guta2020}. While strange from the conventional wisdom of maximum likelihood and Bayesian mean estimation, nonphysical states are actually beneficial for classical shadows.


\textit{Observable-Oriented Pseudo-Likelihood.} Deriving a positive semidefinite model emulating classical shadows remains an intriguing question, however, to eliminate unphysical estimates while retaining the favorable scaling features. With projecting directly onto $\rho_s$ proving unfruitful to this end, we note that, indeed, $\rho_s$ was never intended to serve as an accurate substitute for the true $\rho_g$; instead it facilitates estimates of observables~\cite{Huang2020}. Accordingly, we propose the ``observable-oriented pseudo-likelihood'' 
\begin{equation}
\label{eq:PL2}
L_\bbD(\bx) = \exp\left(-\frac{K}{2} \sum_{n=0}^{N-1} \left|\Tr\rho(\bx)\Lambda_n - \lambda_n^{(s)}\right|^2 \right),
\end{equation}
where we insert the estimates $\lambda_n^{(s)}$ of $N$ observables from $\rho_s$. This formalism ensures only physical values are returned [through the prior $\pi_0(\bx)$], and rates the fitness of proposed states through their overlap with respect to shadow's predictions of observables only. For dimension-independence, we again set $K=MD$ and perform BME for all simulated datasets and $N=3$ above, thinning to $T=2^{10}$ ($T=2^{13}$) for $D=32$ ($D=256$).

The results follow in Fig.~\ref{fig4}. Now BME shows very similar behavior to shadow: the MSE with respect to the ground truth matches shadow results from Fig.~\ref{fig1} closely, though BME still outperforms for $\lambda_2$. Yet unlike shadow, BME here always gives physically permissible estimates ($\lambda_n^{(B)}\in[0,1]$). This pseudo-likelihood therefore attains the goal of a BME model commensurate with classical shadows.

\begin{figure}[tb!]
\centering\includegraphics[width=\columnwidth]{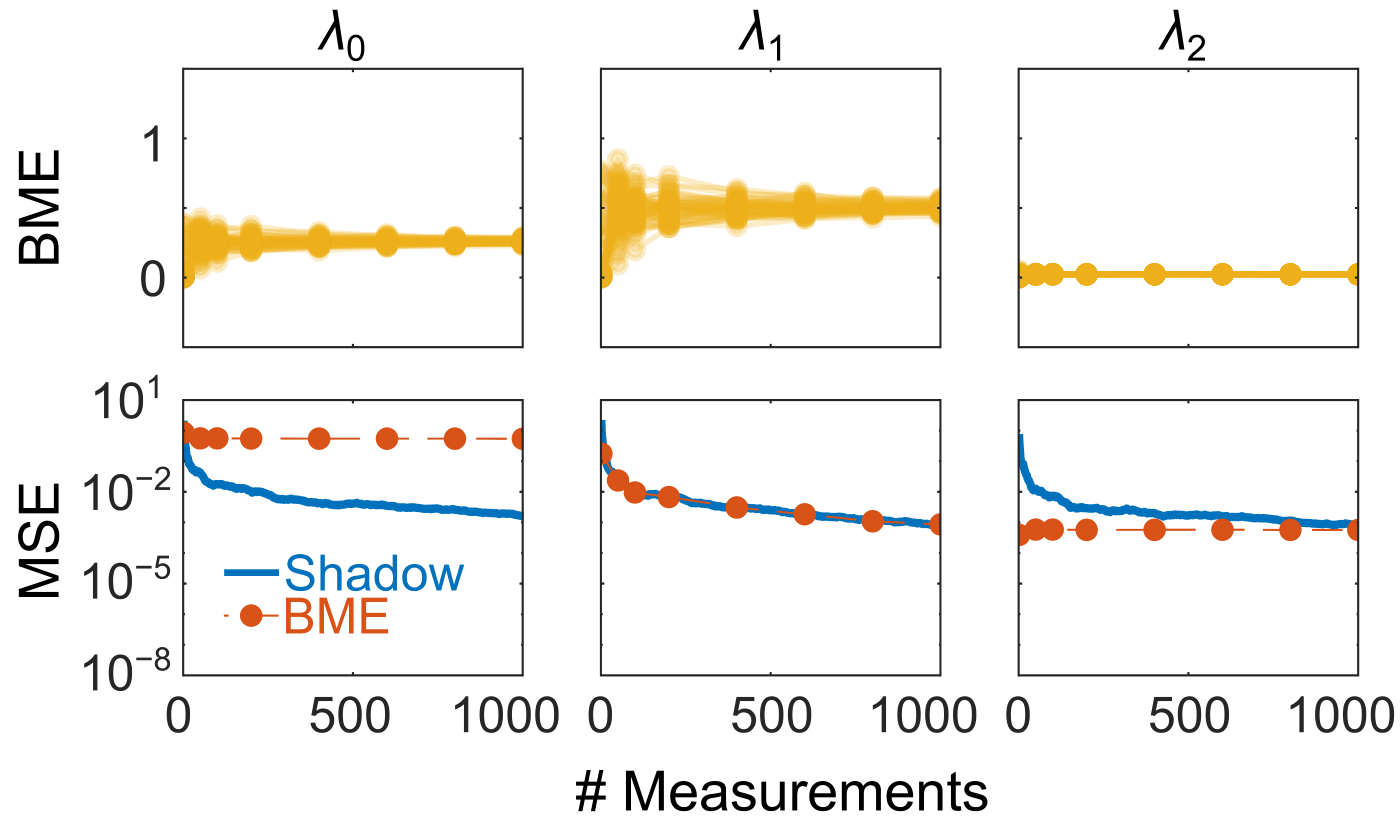}
\caption{Bayesian inference results employing the psuedo-likelihood in Eq.~(\ref{eq:PL3}), for $D=32$. The shadow MSE values from Fig.~\ref{fig1} are reprinted for clarity.}
\label{fig5}
\end{figure}

Yet it is important to emphasize that this approach depends heavily on the quality of the classical shadow. It refines estimates from the shadow with its positive semidefinite requirement, but it does not do markedly better at estimating the ground truth state---at least for arbitrary observables. As an example, we repeat the inference procedure for an observable-oriented pseudo-likelihood based solely on $\Lambda_1$, i.e.,
\begin{equation}
\label{eq:PL3}
L_\bbD(\bx) = \exp\left(-\frac{K}{2} \left|\Tr\rho(\bx)\Lambda_1 - \lambda_1^{(s)}\right|^2 \right),
\end{equation}
which has ground truth value $\lambda_2^{(g)}=\frac{1}{2}$. Results for the $D=32$ case appear in Fig.~\ref{fig5}, where we plot the Bayesian estimates for all three observables even though the psuedo-likelihood is based on $\lambda_1$ only. The estimate $\lambda_1^{(B)}$ closely matches shadow as designed, and $\lambda_2^{(B)}$ agrees with the ground truth well, due to the fact its value is highly probable for a uniform prior. But $\lambda_0^{(B)}\rightarrow \sim\frac{1}{4}$, far from $\lambda_0^{(g)}=1$.

When using the pseudo-likelihood above, all quantum states with identical overlap to $\Lambda_1$ are equally probable, of which the ground truth $\rho_g$ represents just one possibility. The estimate of $\lambda_0$ given only $\lambda_1$ information reflects the inherent uncertainty within this specification. 
So to summarize, our observable-oriented pseudo-likelihood builds physicality into shadow, yet it can only (in general) accurately predict the $N$ observables injected into it: to infer quantities beyond these $N$ can prove unreliable.

\section*{Discussion}
\label{sec:disc}
Our numerical investigations here have elucidated two fascinating features of classical shadows:
\begin{enumerate}
\item Classical shadows perform extremely well at predicting ``unlikely'' observables, i.e., those which obtain high values only on a restricted subset of states within the complete Hilbert space.
\item The nonphysicality of classical shadows is critical to their dimension-independence and accuracy under few measurements.
\end{enumerate}
These findings do not contradict the optimality of Bayesian methods expressed in Eq.~(\ref{eq:opt}):  BME with a full likelihood minimizes MSE for any number and collection of measurements, provided the prior distribution accurately reflects the true knowledge involved. The predictive power of $\rho_s$, then, derives from the fact that the situations in which it is much more accurate that BME are often of particular interest in practice, such as verification of a high-fidelity or highly entangled quantum state. Desiring to extend these features in the Bayesian context, we proposed an observable-oriented pseudo-likelihood that attains shadow's dimension-independence and state-specialized accuracy, with the advantage of guaranteed physicality.

Nonetheless, in all these explorations there remains one prominent sense in which classical shadows unquestionably eclipse BME: computational efficiency. The shadow estimator $\rho_s$ is formed directly from measurements for any dimension $D$; yet computing $\rho_B$ requires tedious MCMC methods, with the number of parameters increasing linearly (quadratically) with $D$ for a pure (mixed) state prior. Here we considered up to $D=256$, a far cry from the $D=2^{120}$ example in Ref.~\cite{Huang2020}, where there is no hope for BME with a parameterization such as ours. Moving forward, it would therefore seem profitable to explore simplified Bayesian models that maintain a fixed parameter dimensionality even as the Hilbert space grows exponentially. For example, if one could specify a prior and likelihood on an observable  $\lambda$ only, to the effect of $\pi(\lambda)\propto L_\bbD(\lambda)\pi_0(\lambda)$, the inference procedure would not be limited directly by exponentially large Hilbert spaces. In this way, Bayesian methods could be extendable to the types of quantum systems sought for practically useful quantum computation.

Overall, our analyses have revealed the value of BME as a tool for shedding light on estimation procedures which formally have no connection to the Bayesian paradigm. The numerical simulations here reveal the complementary strengths of classical shadow and Bayesian tomographic approaches in the efficient estimation of quantum properties. And so we expect valuable opportunities for both methods as quantum information processing resources continue to mature in size and complexity.

\section*{Methods}
\subsection*{Data Simulation Approach}
The method of classical shadows introduced in Ref.~\cite{Huang2020} involves application of a Haar-random (or effectively Haar-random) unitary $U$ followed by measurement in the computational basis. We exploit the fact that our target state is pure to substantially reduce the complexity of simulating this procedure. In particular, our simulation method requires the generation of only size-$D$ random vectors rather than $D\times D$ random unitaries.

Without loss of generality we work in a rotated basis such that the first basis state coincides with the ground truth: $\rho_g = \ket{0}\bra{0}$. Then the probability of observing outcome $j$ depends only on $|\braket{j|U|0}|^2 = |U_{j0}|^2 = |(U^\dagger)_{0j}|^2$. That is, the distribution of outcomes depends only on the first row of $U^\dagger$. Now, when $U$ is Haar-random, each individual row and column of $U^\dagger$ is a uniformly distributed length-1 vector $u$. Furthermore, given any component $u_j$, the remaining components are a uniformly distributed vector of length $\sqrt{1-|u_j|^2}$. A uniformly random vector $u$, corresponding to the first row of $U^\dagger$, may be obtained by generating $D$ complex normal random values and normalizing them to yield a unit length vector. An outcome $n\in \{0,1,\ldots,D-1\}$ is then chosen with probability $|u_n|^2$. This selects the $n$th column of $U^\dagger$. Since this column (whichever it is) is uniformly distributed, its remaining elements are uniformly distributed with length $\sqrt{1-|u_n|^2}$. The explicit procedure is as follows:

\begin{enumerate}
\item Posit a measurement unitary $U_m^\dagger = [\tilde{\varphi}_0 \cdots \tilde{\varphi}_{D-1}]$, where each $\tilde{\varphi}_n$ is a column vector corresponding to one of the $D$ possible output states.
\item Generate $D$ complex normal samples $w_n \stackrel{\textrm{i.i.d.}}{\sim}\cN(0,1) + i\cN(0,1)$ and normalize
\begin{equation}
\label{eq:row}
u_n = \frac{w_n}{\sqrt{\sum
\limits_{n^\prime=0}^{D-1} |w_{n^\prime}|^2}}.
\end{equation}
These define projections of the unitary's basis states on the ground truth: $u_n = \braket{0|\tilde{\varphi}_n}$, or in other words, the elements in the first row of $U_m^\dagger$.
\item Select an integer $n\in\{0,1,...,D-1\}$ at random with probability $|u_n|^2$. This implies that the state $\tilde{\varphi}_n$ is detected.
\item Generate $D-1$ complex normal samples $v_j \stackrel{\textrm{i.i.d.}}{\sim}\cN(0,1) + i\cN(0,1)$ ($j=1,2,...,D-1$). These set the remaining coefficients of the detected state $\tilde{\varphi}_n$.
\item Finally, take
\begin{equation}
\label{eq:col}
\ket{\psi_m} = 
u_n\ket{0} +  \sqrt{ \frac{1-|u_n|^2} {\sum\limits_{j^\prime=1}^{D-1} |v_{j^\prime}|^2}} \sum_{j=1}^{D-1} v_j  \ket{j}
\end{equation}
as the measured state. 
\end{enumerate}


Utilizing this method, we performed 50 independent trials with 1000 measurements each, for Hilbert space dimensions $D=32$ and $D=256$, giving a total of 100 datasets which are used in all subsequent tests above. The two values of $D$ were selected specifically to clarify how classical shadows and BME differ in their scaling with dimension.

\vspace{-0.1in}
\section*{Acknowledgments}
\vspace{-0.15in}
This work was funded by the U.S. Department of Energy, Office of Advanced Scientific Computing Research, through the Quantum Algorithm Teams and Early Career Research Programs. This work was performed in part at Oak Ridge National Laboratory, operated by UT-Battelle for the U.S. Department of Energy under contract no. DE-AC05-00OR22725.

\end{document}